\RequirePackage{lineno}
\documentclass[aps,amsfonts,amssymb,nofootinbib,twocolumn,amsmath,superscriptaddresssquare]{revtex4-2}
\usepackage{graphics}
\usepackage{hyperref}  
\usepackage{color}   
  
\usepackage{hyperref}
\hypersetup{
  colorlinks,
  citecolor=magenta,
  linkcolor=blue,
  urlcolor=blue}
   
\usepackage{graphicx}% Include figure files
\usepackage{bm}% bold math
\usepackage{hyperref}
\usepackage{natbib}
 
\begin{document}
%\linenumbers
   
\title{Planckian scattering and parallel conduction channels in the iron chalcogenide superconductors FeTe$_{1-x}$Se$_x$}

\author{Ralph Romero III}
\email{rromero@jhu.edu}
\affiliation{William H. Miller III Department of Physics and Astronomy, The Johns Hopkins University, Baltimore, Maryland 21218, USA}

\author{Hee Taek Yi}
\affiliation{Department of Physics and Astronomy, Rutgers: the State University of New Jersey, Piscataway, NJ, 08854,  USA}

\author{Seongshik Oh}
\affiliation{Department of Physics and Astronomy, Rutgers: the State University of New Jersey, Piscataway, NJ, 08854,  USA}

\author{N. P. Armitage}
\email{npa@jhu.edu}
\affiliation{William H. Miller III Department of Physics and Astronomy, The Johns Hopkins University, Baltimore, Maryland 21218, USA}

\date{\today}

\maketitle

\textbf{The remarkable linear in temperature resistivity of the cuprate superconductors, which extends in some samples from $T_c$ to the melting temperature, remains unexplained. Although seemingly simple, this temperature dependence is incompatible with the conventional theory of metals that dictates that the scattering rate, $1/\tau$, should be quadratic in temperature if electron-electron scattering dominates. Understanding the origin of this temperature dependence and its connection to superconductivity may provide the key to pick the lock of high-temperature superconductivity. Using time-domain terahertz spectroscopy (TDTS) we elucidate the low temperature conducting behavior of two FeTe$_{1-x}$Se$_x$ (FTS) samples, one with almost equal amounts of Se and Te that is believed to be a topological superconductor, and one that is more overdoped.  Constrained with DC resistivity, we find two conduction channels that add in parallel, a broad one in frequency with weak temperature dependence and a sharper one whose scattering rate goes as the Planckian limited rate, $\sim kT/h$. Through analysis of its spectral weight we show the superconducting condensate is mainly drawn from the channel that undergoes this Planckian scattering. }

\bigskip

The chalcogenide superconductor series FeTe$_{1-x}$Se$_x$ (FTS) provides a powerful platform to explore strongly correlated superconductivity in proximity to other ordered states~\cite{DiscoveryofSCinFeSe,her_anisotropy_2015}. The end member FeTe does not superconduct and possesses a tetragonal to monoclinic structural transition accompanied by an antiferromagnetic (AFM) transition around 70K~\cite{FeTeAFM1,FeTeAFM2}. The other end member FeSe undergoes a tetragonal to orthorhombic structural transition into a nematic phase at 90K and superconducts at 9 K~\cite{sym12091402}. At an ``optimal" doping, i.e. close to equal parts Se/Te ($x = 0.45$), $T_c$ is elevated to 14.5 K with upper critical fields reaching 50T~\cite{her_anisotropy_2015}. Along with this maximized $T_c$, samples with optimal doping have amassed a compelling body of evidence~\cite{yin_observation_2015,zhang_observation_2018,wang_evidence_2018,machida_zero-energy_2019,TRSARPES,TRSsagnac,roppongi2025topologymeetstimereversalsymmetry} suggesting FTS belongs to a class of topological superconductors able to host Majorana zero modes (MZMs), that are proposed to be a platform for topological quantum computation~\cite{sato_topological_2017}.

\begin{figure*}[t]
    \includegraphics[]{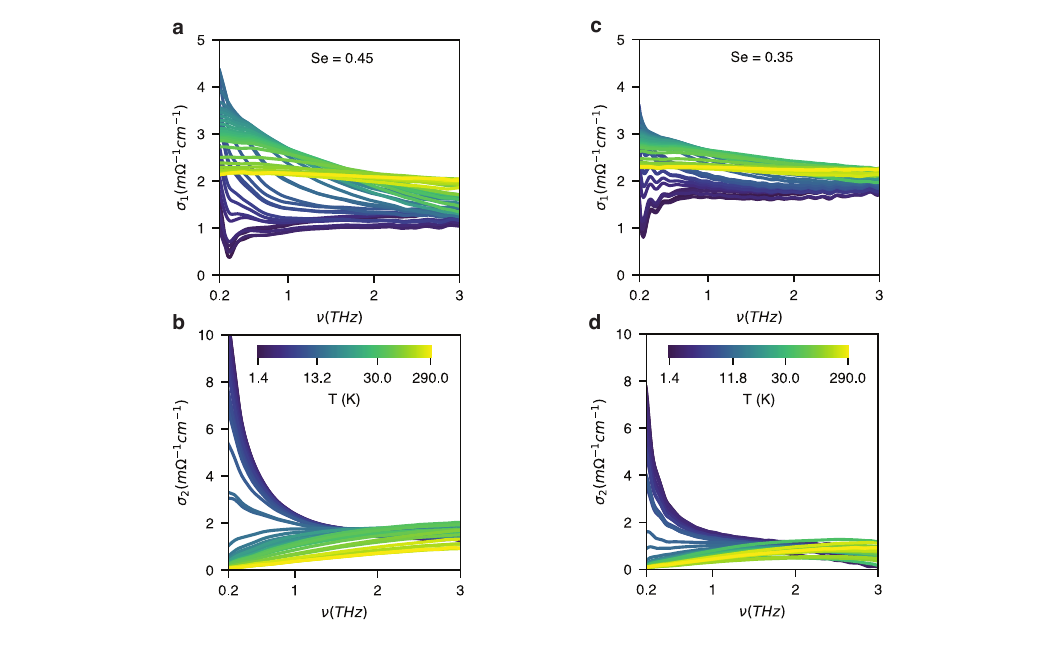}
    \caption{\textbf{THz conductivity of the Se = 0.45 and 0.35 samples.} Real and imaginary parts of the THz conductivity from room temperature down to 1.4K measured from 0.2 to 3 THz for the $x = 0.45$ sample a), b) and the $x = 0.35$ sample c), d).}
    \label{fig:fig1}
\end{figure*}

The iron chalcogenide superconductors have a phenomenology reminiscent of the cuprate superconductors, namely a superconducting dome whose peak straddles a quantum critical point (QCP)~\cite{mukasa_enhanced_2023}. While this is interesting, it should come as little surprise as out of the iron based superconductors and other pnictides, FeTe and FeSe have the highest relative mass enhancements~\cite{yin_kinetic_2011}, suggesting correlations play a large role in their behavior.  In retrospect it is also not surprising that a combination of these correlated parent compounds gives birth to unconventional superconductivity, but what is so special about the optimal ($x = 0.45$) doping that gives rise to an increased transition temperature and ultimately such a unique superconducting state? In the cuprates it is often proposed that $T_c$ is maximized at the doping that coincides with a zero temperature phase transition. It has been conjectured that proximity to this QCP may give rise to strong quantum fluctuations, which manifest themselves as deviations from the Fermi liquid picture, namely resistivity or equivalently the charge scattering rate that goes as as Planckian rate, $~\sim kT/h$ \cite{Legros2019} (and linear in frequency single particle self-energies).  Such a scenario presents challenges however.  In FTS, it was initially conjectured that critical fluctuations stemmed from an AFM QCP associated with the transition in FeTe~\cite{homes_optical_2011}. However, recently it was shown that optimal doping for FTS occurs near a QCP associated with a nematic order~\cite{mukasa_enhanced_2023} and theoretically a clean nematic QCP is expected to give a imaginary self-energy that goes like $\omega^{2/3}$~\cite{guo2022large}.

A time-resolved technique such as TDTS allows the experimenter to measure the magnitude and phase of the transmitted electric field simultaneously. This allows one to measure both parts of the complex conductivity in the THz range, $\sigma_1(\nu)$ and $\sigma_2(\nu)$, without having to use the Kramers-Kronig relations. TDTS was performed on two 100 nm FeTe$_{1-x}$Se$_x$ films with $x = 0.45$ and $0.35$ grown by molecular beam epitaxy on 0.5 mm thick CaF$_2$ substrates, capped with 100 nm of Se. The two samples have similar transition temperatures of 13.2 and 11.8 K, respectively, measured via DC resistivity (Extended Data Fig.~\ref{Extended Data:ExDat1}). Fig.~\ref{fig:fig1} shows $\sigma_1(\nu)$ (top) and $\sigma_2(\nu)$ (bottom) for the two samples. The left side shows the complex conductivity for the optimally doped sample, while the right side corresponds to the $x = 0.35$ sample. Both samples show the same qualitative behavior. At high temperatures, $T \gg T_c$, $\sigma_1(\nu)$ is broad and featureless while $\sigma_2(\nu)$ is small and slowly increasing with frequency. This is canonical behavior of a metal whose scattering rate is larger than the measurement window (i.e. $\Gamma \gg$ 3 THz). As temperature decreases $\sigma_1(\nu)$ begins to increase and acquire negative slope reminiscent of a Drude peak. Once the superconducting transition temperature is crossed, $\sigma_1(\nu)$ decreases while $\sigma_2(\nu)$ begins to show $1/\nu$ dependence, corresponding to the onset of superconductivity as spectral weight moves below our measurement range into the delta function at $\nu = 0$. In a conventional scenario one would expect most of the spectral weight below the gap energy, $2\Delta = 3.5 k T_c  \approx 1$ THz, to condense to $\nu = 0$, much like what has been seen in disordered NbN thin films~\cite{BingNbN}. However, in the present case even at the lowest temperature T = 1.4 K, there is still considerable spectral weight within the gap.
 \begin{figure*}[t]
\includegraphics[width = \textwidth]{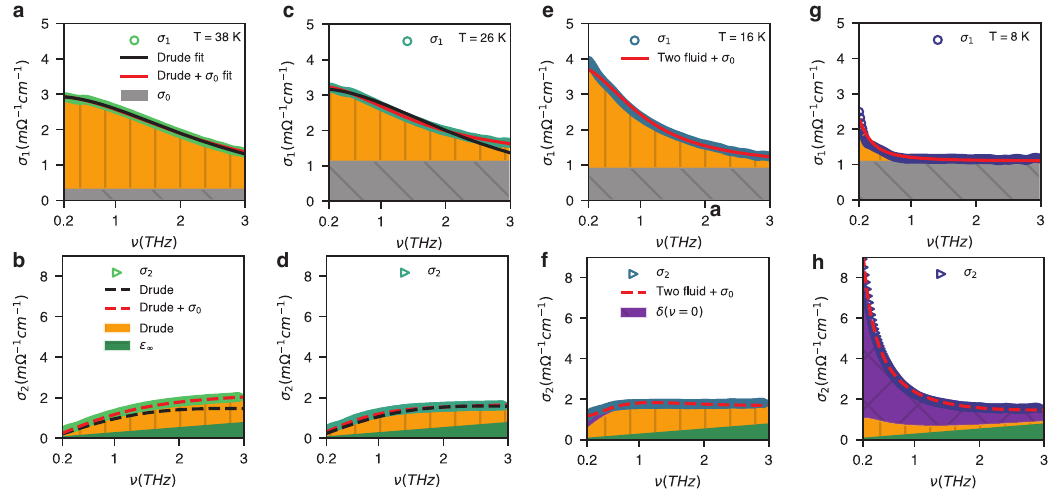}
\caption{\textbf{Fits to the THz conductivity of the Se = 0.45 sample at various temperatures.} a), b) $\sigma_1$, $\sigma_2$ at a temperature (38 K) where a single Drude term suffices. c), d) A temperature (26 K) where a constant offset, $\sigma_0$, must be included alongside the Drude term. e), f) A temperature, above $T_c$, (16 K) where the two fluid model must be invoked. g), h) The lowest temperatures where the conductivity becomes mainly sensitive to the condensate at $\nu = 0$. }
    \label{fig:fig2}
\end{figure*}

\begin{figure*}[t]
	\includegraphics[]{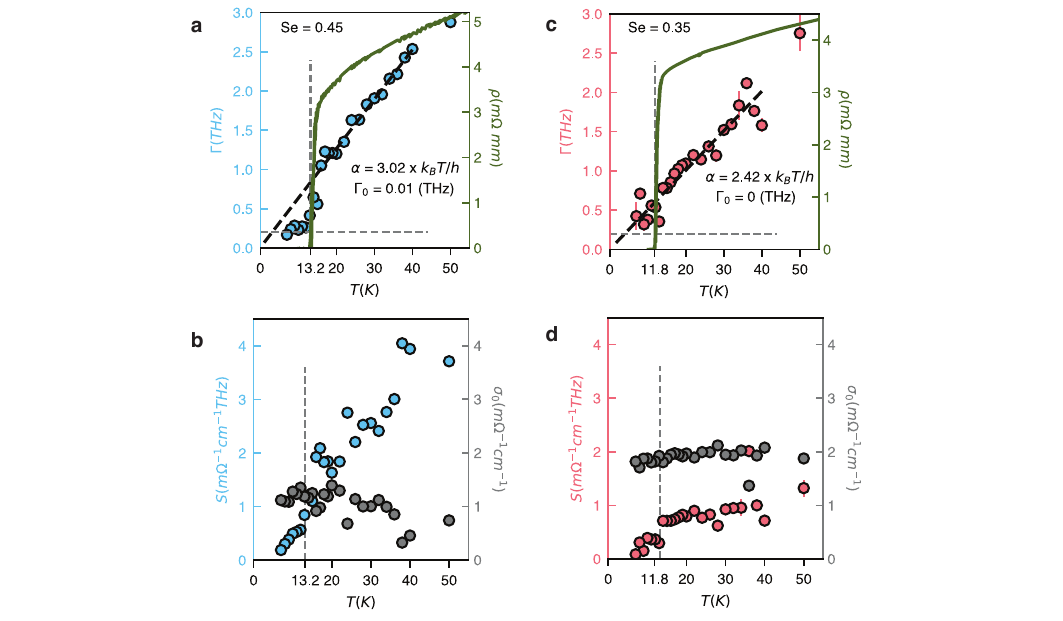}
    \centering
	\caption{\textbf{Fit parameters along with DC resistivity.}  The left column correspond to the $x= 0.45$ sample, while the right to $x=0.35$ sample. a) and c) show the scattering rate, $\Gamma$, with the vertical dashed line indicating $T_c$ and the horizontal dashed line denoting the edge of our spectral range. Linear in temperature scaling can be observed with slopes on the order of $3  k T/h$. b) and d) Show the spectral weight of the Drude term, $S$, (left axis) and the constant offset, $\sigma_0$, (right axis).}
	\label{fig:fig3}\
\end{figure*}

\begin{figure*}[t]
	\includegraphics[]{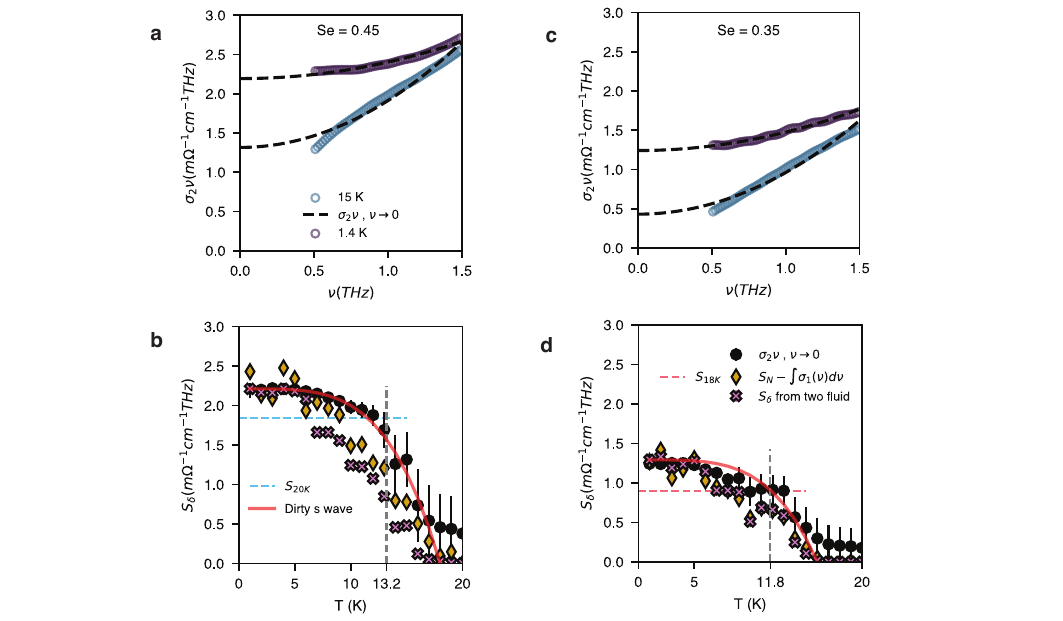}
    \centering
	\caption{\textbf{Superfluid density, three ways.} The superfluid density measured using three methods for both samples. The left column corresponds to the optimally doped sample, while the right to $x=0.35$ sample. a) and c) show fits then extrapolations of $\sigma_2 \nu$ to $\nu = 0$. At temperatures around $T_c$ the quadratic form is incorrect. b) and d) show the comparison of the three methods. The purple x's indicate $S_{\delta}$. The goldenrod diamonds show the value expected in the context of single band sum rules.  The black circles are the results of a) and c). The vertical dashed lines indicate $T_c$ while the horizontal dashed lines indicate the spectral weight of the Drude peak, $S$, at 2K above the onset of superconducting fluctuations.}
	\label{fig:fig4}\
\end{figure*}
At high temperatures, the spectra look like a simple Drude oscillator with a scattering rate on the order of a few THz that goes superconducting at low temperature.  However, as the temperature is lowered it becomes clear that there is more structure.  We can understand the spectra from the perspective of a model with a Drude-like term, a constant in frequency term, and a superconducting contribution (as well as the $\epsilon_{\infty}$ term from the polarizability of the lattice), $\tilde\sigma(\nu) = S/(\Gamma - i\nu) \ + \ \sigma_{0} \ + \ \frac{\pi}{2}S_{\delta}\delta(\nu=0) + \ iS_{\delta}/\nu \ - \ i \epsilon_0(\epsilon_{\infty} - 1)\nu$. Here $S$ is the spectral weight of the normal state carriers, $\Gamma$ is the scattering rate for normal electron-like carriers, $\sigma_{0}$ is a temperature dependent residual conductivity, $S_{\delta}$ is the spectral weight of the superfluid density at $\nu =0$, and $\epsilon_{\infty}$ parametrizes the contribution from the lattice polarizability. In accord with previous optical conductivity results \cite{homes_optical_2011} we set $\epsilon_{\infty} = 4 $ in our fits. At high temperatures, Figs.~\ref{fig:fig2}a) and b), the data is well fit by a single Drude oscillator which is suggestive of interband scattering that dominates over intraband scattering, which may blur the multiband picture typically associated with FTS. As the temperature decreases, one can see in Fig.~\ref{fig:fig2}c) and d) that the single Drude picture begins to fail and a constant offset term, $\sigma_0$, must be included. This offset can be thought of a another Drude contribution, but one with a scattering rate much larger than our spectral range. At these intermediate temperatures intraband scattering becomes dominant and the multiple contributions can be resolved.  As we approach $T_c$, Fig.~\ref{fig:fig2}e) and f), the $1/\nu$ term must be invoked at temperatures slightly above the transition, indicative of superconducting fluctuations. Finally, as we reach the lowest temperatures, the single Drude peak continues to narrow, ultimately moving below our spectral range, leaving us with spectra that are dominated by the imaginary part of superfluid term and the $\sigma_0$ contribution to the real part.  Extended Data Fig.~\ref{Extended Data:ExDat2} shows similar fits to the $x = 0.35$ data.

Fig.~\ref{fig:fig3} shows the results of the fit parameters for both samples for intermediate temperatures where both the Drude and $\sigma_0$ terms are significant.  On the top row we show the scattering rate $\Gamma$ alongside the DC resistivity, $\rho$, for the $x = 0.45$ a) and $x = 0.35$ c) samples.  $\Gamma$ is plotted from 50 K until the Drude peak becomes so narrow that we cannot measure it ($\Gamma < 0.2$THz).  Notably a clear linear dependence in temperature is evident in $\Gamma$. The rates are fit to $\Gamma_0 + \alpha kT/h$, from 20 K to 40 K and shown extrapolated to T = 0 K (Fit parameters are displayed on their respective graphs).  Both samples display linear-in-T scaling with $\alpha = 3.02 $ and $\Gamma_0=  10$ GHz for the $x = 0.45$ sample and $\alpha = 2.42 $ and $\Gamma_0< 10$ GHz  for the $x = 0.35$ sample. $S$ and the DC value of $\sigma_0$ are shown in the bottom row of Fig.~\ref{fig:fig3}. For the $x = 0.45$ sample, $S$ decreases considerably as temperature goes through $T_c$ seemingly going towards zero, while $\sigma_{0}$ has little if any temperature dependence through the transition. $S$ of the $x = 0.35$ sample displays qualitatively similar behavior as above but on a smaller scale, with the offset having a weak temperature dependence.

This parsing of the THz conductivity reveals a conduction channel with a hidden Planckian dissipation component that is hidden in the distinctly non linear in T resistivity in FTS.  With this in mind, one notes the qualitative temperature dependence of the DC conductivity, $\sigma_{DC} = 1/\rho$, can be recovered when adding $\sigma_0$ to a $\sigma_{Drude}(\nu \rightarrow 0) $ that goes like $S/\Gamma$ (Extended Data Fig.~\ref{Extended Data:ExDat3}). Although this relationship may seem obvious, it has deep implications about the underlying electronic conduction. Matthiessen's rule provides the expectation that for a simple, single band metal, the different scattering channels add linearly e.g. $\Gamma_{Total} = \Gamma_{Intrinsic} +\Gamma_{Impurity}$ + etc. The net effect of different scattering mechanisms can be understood in the same fashion as how different resistors in series combine i.e. resistances are summed. However, our data is incompatible with this picture. We observe not the linear combination of resistive channels, but the linear combination of conduction channels.  From our fits we know $\Gamma$ is linearly proportional to T, thus $\sigma _{DC} = \sigma_0 + S/\Gamma = \sigma_0 + \frac{S}{\alpha k T / h} $ which can be taken to be $ a + b/T$. This gives a DC resistivity that goes as $\rho_{DC} = \frac{1}{a+ b/T}$.  The fits (Extended Data Fig.~\ref{Extended Data:ExDat1}) using this remarkably simple function match the data well from 200K to 25K, which is the temperature superconducting fluctuations begin to appear. 

This perspective of parallel conduction channels has its antecedents~\cite{HusseyResistors, vanHeumenStrangeMetal,Clayhold2010}. Here this idea provides an interpretation of our spectra where there are two distinct conduction channels, a slower relaxing channel which has a scattering rate within our measurement frequency window and has a Planckian temperature dependence, and a fast relaxing channel whose scattering rate is much larger than our spectral window. From Fig.~\ref{fig:fig3}b) and d) it is clear that both channels are affected by the superconducting transition differently. The fast channel appears more or less oblivious to the phase transition, while the spectral weight of the slow channel is suppressed. From Fig.~\ref{fig:fig2}, one notices the low frequency spectral weight is suppressed. Superconductivity is characterized by a $\delta(\nu = 0)$ whose spectral weight equals the spectral weight lost in the finite frequency conductivity. In the case of a dirty limit BCS s-wave superconductor this spectral weight comes from the states below the gap.  As we show quantitatively below, this spectral weight is mainly drawn from of channel that displays the NFL Planckian behavior. 

The nature of the pairing mechanism in FTS remains an open question. Visually our spectra are reminiscent of those from other ``unconventional" superconductors such as LCCO~\cite{ZhenisdWave}, LSCO~\cite{FahadMissingSpectral}, and Sr$_2$RuO$_4$~\cite{YouchengSRO214}.   We can quantify the T dependence of the superfluid density $S_{\delta}$ from the two fluid model fits, plotted in Fig.~\ref{fig:fig4} b) and d) as the purple x's.  Another approach to determining $S_{\delta}$ is via the expected low frequency functional form for $\sigma_2(\nu) \propto (S/\Gamma^2) \nu \ + \ S_{\delta} / \nu $. By fitting $\sigma_2 \nu$ with a quadratic form, then extrapolating to $\nu = 0$ we can obtain $S_{\delta}$~\cite{ZhenisdWave}. These fits are shown in the top row of Fig.~\ref{fig:fig4}. At the lowest temperatures the quadratic form performs well as the superconducting state dominates our spectra. For temperatures closer to $T_c$ where this form is expected to be less valid, the parabolic shape fits not as well as seen in  Fig.~\ref{fig:fig4} a) and c). The temperature dependence of the superfluid density from these fits is shown in Fig.~\ref{fig:fig4} b) and d) as the black dots.

In order to accurately capture the superfluid density at temperatures closer to the transition, we appeal to the Ferrel-Glover-Tinkham (FGT) optical sum rule~\cite{FGT1,FGT2} which is $ S_{\delta} = S_N - S_U$, where $S_U$ is the spectral weight of uncondensed carriers in the superconducting state and  $S_N$ is the normal state spectral weight at $T_{c0} + 2K$.  The FGT sum rule allows one to compute $S_{\delta}$ as the difference in area between the normal state conductivity $S_N$ and the conductivity at some temperature below $T_c$, $S_U =\frac{\pi}{2}\int_0^\infty \sigma_1(\nu) d\nu$. $S_{\delta}$ computed from the FGT sum rule is plotted as the goldenrod diamonds in Figures 4b) and d). 

At temperatures above $T_c$, $S_{\delta}$ is still non-zero for both samples, indicative of either spatial or temporal fluctuations of the condensate. Of course, the delta function is lost outside the superconducting state, and so the present observation is consistent with an ``almost" condensed spectral weight that has a narrow but finite width at temperatures near $T_c$ in the normal state.  For the optimally doped sample shown in Fig.~\ref{fig:fig4}b), once $S_{\delta}$ becomes finite there is a monotonic rise until it saturates at around 7K. The $x = 0.35$ sample, Fig.~\ref{fig:fig4}d), shows similar behavior, albeit on a smaller scale.  It is interesting to note that ultimately the $x = 0.35$ has considerably less spectral weight at zero temperature relative to the optimally doped sample given their similar transition temperatures.

The temperature dependence of the superfluid density is at first glance rather unremarkable from the perspective of ``unconventional" superconductivity. We overlay the generic form for the superfluid density of a dirty s-wave superconductor~\cite{tinkham} as a guide to eye. However to obtain reasonable fits, the fitting parameter $T_{c0}$ for the $x = 0.45$ sample needed to be 18K and for the $x = 0.45$, $T_{c0} =$  16K.  This can be interpreted as either a larger than BCS-sized low temperature gap ($\frac{2 \Delta_{BCS}}{k T_c} = 3.53$) or a temperature dependence that differs markedly from that of BCS (particularly near $T_c$). This may be expected for a strongly fluctuating superconductor. 

In Fig.~\ref{fig:fig4} b) and d), the spectral weight at $T_{c0} + 2K$ of the Drude peak is plotted as the horizontal dashed line. For both samples we find that the low temperature value of $S_{\delta}$ just overshoots this value of the spectral weight. This indicates that the charge carriers that undergo Planckian scattering are the principal participants in the condensate. The spectral weight in the incoherent background that gives $\sigma_0$ does not contribute to superconductivity.

The origin of $\sigma_0$ and its significance as a low temperature in-gap spectral weight is interesting. It is in qualitative agreement with the optical conductivity computed in Ref.~\cite{islam2024} in the context of superconductivity mediated by nematic fluctuations (NFMS) where low lying excitations near cold spots provide in gap states at low temperature. The presence of in-gap states have also been observed in other superconductors with an anisotropic order parameter~\cite{FahadMissingSpectral, ZhenisdWave}, albeit there they appear in a different form.  As predicted by~\cite{islam2024}, for NFMS in doped FeSe with impurity scattering, the temperature dependence of the superfluid density resembles that of a dirty BCS SC, with a $T\rightarrow0$ limit of $S_{\delta}$ being half of the normal state spectral weight or less with increasing impurity scattering ($S_{\delta}/S_{N} \leq 0.5$). We can compare to this prediction with $S_N$ being the area under $\sigma_1$ at $T_{c0} + 2K$ and $S_{\delta}$ being the average of the three methods at 1.4K. For the optimally doped sample $S_{\delta}/S_{N} = 0.53$ and the $x = 0.35$ sample $S_{\delta}/S_{N} = 0.30$. Both of these values agree with the predictions of NFMS in the presence of impurities.  We would also point out that the superfluid density appears to fall faster than the s-wave BCS prediction.   This is also in line with the predictions of the NFMS theory~\cite{islam2024}.   However we would caution that in order to get a clear view of its temperature dependence it is essential to measure the superfluid density with a lower frequency probe like transverse field $\mu$SR as done on FeSe$_{1-x}$S$_{x}$ ~\cite{muSRsuperfluidDensity}. 

In conclusion, by performing time domain terahertz spectroscopy on FTS thin films, we observed two distinct conduction channels. One channel which remains with finite conductivity at the lowest temperature has a relaxation rate larger than of our spectral range. We also observe a slow channel which displays non-Fermi liquid-like behavior that exhibits a linear in temperature scattering rate of the Planckian form $\sim 3  k T/h$. Such NFL-like response has been seen in related materials such as FeSe~\cite{kasahara_giant_2016,jiang_interplay_2023} and FeSe$_{1-x}$S$_{x}$~\cite{huang_non-fermi_2020}.  It and the presence of the residual low temperature conduction maybe related to the proximity to a nematic QCP, which occurs near optimal doping. Our observation bears close resemblance to a recent theoretical prediction based on such a scenario~\cite{islam2024}.  Most strikingly, we observe that at low temperature the spectral weight of the condensate matches that of the Planckian dissipated channel in the normal state. This suggests that the correlations that give this anomalous normal state scattering are related to those that cause superconductivity. We hope our work motivates further studies of the normal state of FeTe$_{1-x}$Se$_x$.

\section{Methods: }

FTS thin films used in this study were prepared on 10 × 10 × 0.5 mm$^3$ CaF$_2$ (100) substrates using a custom-built MBE system (SVTA) with a base pressure of ~10$^{-10}$ Torr. CaF$_2$ substrates were in situ cleaned at 400°C for 15 minutes prior to FTS film growth at 300°C. High-purity Fe, Te, and Se sources were thermally evaporated using Knudsen diffusion cells for the film growth. All source fluxes were calibrated in-situ using a quartz crystal microbalance and ex-situ by Rutherford backscattering spectroscopy.  Se capping layers were deposited at room temperature on top of the films.
 
Time domain terahertz spectroscopy was performed in a home built spectrometer utilizing a fiber coupled Toptica TeraFlash Pro emitter and detector. The sample is placed in an open flow liquid helium cryostat capable of reaching 1.4K by filling the cryostat with liquid helium and pumping appropriately. By measuring the transmitted electric field through the sample, $\tilde{E}_{Samp}(t)$, and a bare substrate reference, $\tilde{E}_{Ref}(t)$, and taking their Fourier transforms, one can apply the Fresnel equations then exploit the thin film approximation to compute the complex conductivity $\tilde{\sigma}(\omega)$ directly via the relation, 
\begin{equation}
    \tilde{\sigma}(\omega)=\frac{n+1}{dZ_0} \left ( \frac{\exp(i\frac{\omega}{c}\Delta L (n-1))}{\tilde{T}(\omega)}-1 \right), 
\end{equation}
where $\tilde{T}(\omega) = \frac{\tilde{E}_{Samp}(\omega)}{\tilde{E}_{Ref}(\omega)}$, $n$ is the index of the substrate, $Z_0$ is the impedance of free space, $d$ is the thickness of the thin film and $\Delta L$ is the thickness difference between the the substrate the sample was grown on and the substrate used as reference.

\section{Acknowledgments: }
 
This work at JHU and Rutgers was supported by the ARO MURI ``Implementation of axion electrodynamics in topological films and device" W911NF2020166. The instrumentation development at JHU that made these measurements possible was supported by the Gordon and Betty Moore Foundation EPiQS Initiative Grant GBMF-9454. We would like to thank A. Chubukov and P. Coleman for encouraging discussions. 

\bibliography{main}

\setcounter{figure}{0}
\renewcommand{\figurename}{Extended Data Fig.}

\begin{figure*}[t]
	\includegraphics[]{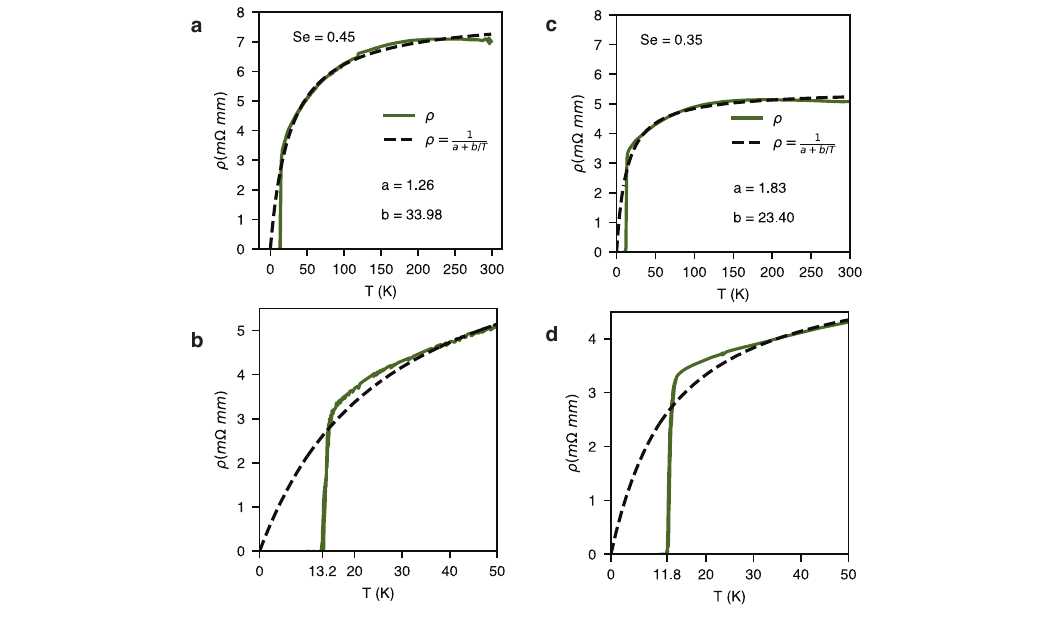}
    \centering
	\caption{\textbf{DC resistivity from transport} $\rho_{DC}$ of both samples fit with the functional form derived in the main text. The top row shows the full temperature range while the bottom is zoomed in to 50 K. a), b) correspond to the $x = 0.45$ sample and c), d) to the $x = 0.35$ sample}
	\label{Extended Data:ExDat1}\
\end{figure*}

\begin{figure*}[h]
	\includegraphics[]{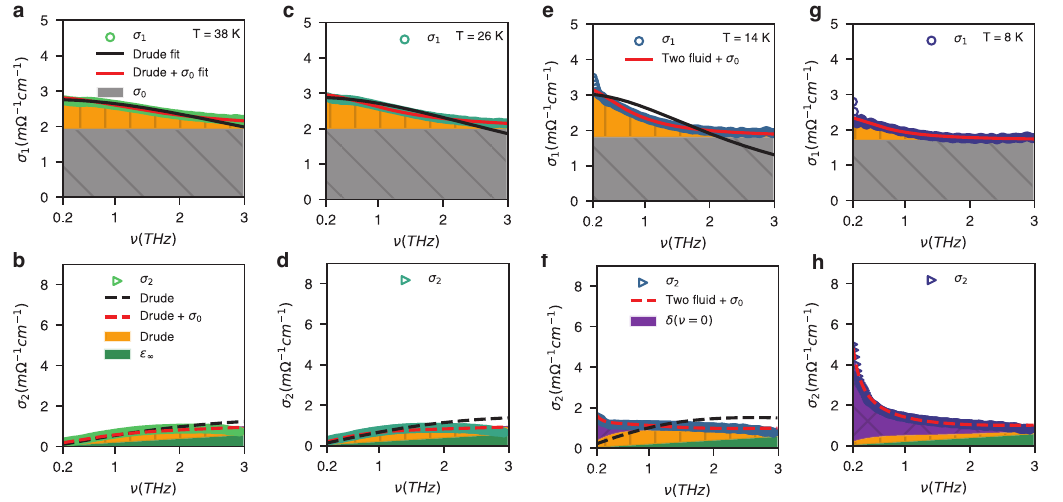}
    \centering
	\caption{\textbf{Fits to the THz conductivity of the Se = 0.35 sample.} Fig. ~\ref{fig:fig2} from the main text for the $x = 0.35$ sample}
	\label{Extended Data:ExDat2}\
\end{figure*}

\begin{figure*}[h]
	\includegraphics[]{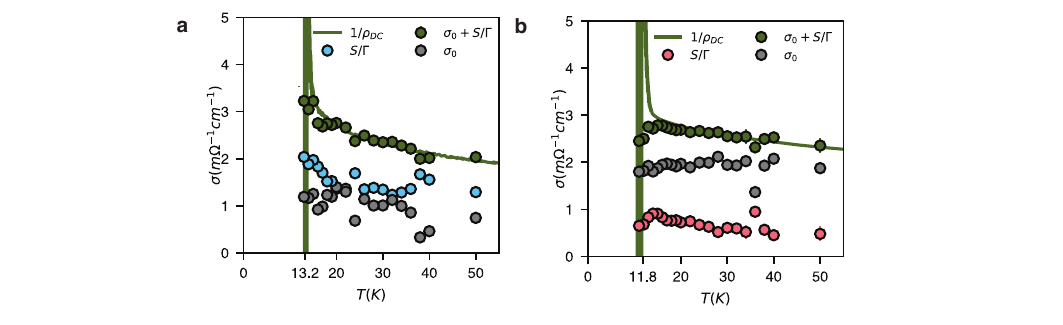}
    \centering
	\caption{\textbf{Hint of parallel conduction channels.} Experimental motivation for the starting point to derive functional form which fits DC data. a) corresponds to the $x = 0.45$ sample and b) to the $x = 0.35$ sample}
	\label{Extended Data:ExDat3}\
\end{figure*}

\end{document}